%% file: main.tex
\documentclass[journal]{IEEEtran}

\usepackage[T1]{fontenc}
\usepackage[utf8]{inputenc}
\usepackage{amsmath,amssymb,amsfonts}
\usepackage{graphicx}
\usepackage{textcomp}
\usepackage{xcolor}
\usepackage{booktabs}
\usepackage{array}
\usepackage{tabularx}
\usepackage{url}
\usepackage{hyperref}
\usepackage{cite}
\usepackage{balance}
\usepackage{microtype}
\usepackage{float}
\usepackage{caption}

\hypersetup{
  colorlinks = true,
  linkcolor  = black,
  citecolor  = black,
  urlcolor   = blue
}

\newcolumntype{P}[1]{>{\centering\arraybackslash}p{#1}}
\newcolumntype{L}[1]{>{\raggedright\arraybackslash}p{#1}}

\begin{document}

\markboth{Preprint --- Submitted to \emph{Internet of Things} (Elsevier)}%
{More \MakeLowercase{\textit{et al.}}: Multi-Interface Firmware Acquisition and Validation for Consumer Quadcopters}

\title{A Multi-Interface Firmware Acquisition and Validation Methodology
for Low-Cost Consumer Drones:
A Case Study on Three Holy Stone Platforms}

\author{%
  Sandesh More,
  Sneha Sudhakaran, and
  Marco Carvalho%
  \thanks{S. More and S. Sudhakaran are with the Department of Computer
    Science, Florida Institute of Technology, Melbourne, FL 32901, USA
    (e-mail: smore2022@my.fit.edu; ssudhakaran@fit.edu).}%
  \thanks{M. Carvalho is with the Department of Electrical Engineering
    and Computer Science, Florida Institute of Technology, Melbourne,
    FL 32901, USA (e-mail: mcarvalho@fit.edu).}%
  \thanks{\textbf{Preprint notice.} This manuscript is a preprint version
    of a paper that is being prepared for submission to the journal
    \emph{Internet of Things} (Elsevier, ISSN 2542-6605,
    \url{https://www.sciencedirect.com/journal/internet-of-things}).
    A revised version may differ from this preprint following peer review.}%
  \thanks{Corresponding author: Sandesh More.}%
}

\maketitle

\begin{abstract}
Consumer unmanned aerial vehicles (UAVs) have evolved into capable computing platforms,
yet their embedded firmware remains largely inaccessible to the security community.
Entry-level models, in particular those marketed to first-time and younger operators,
commonly ship with limited protection mechanisms and no public documentation of their
software internals.
This paper presents a systematic study of firmware extraction and validation applied to three
Holy Stone consumer drone models: the HS175D, HS720, and HS360S.
Rather than pursuing reverse-engineering outcomes, the work focuses on obtaining reliable,
ground-truth firmware images across heterogeneous hardware designs using only
commercially available, low-cost tooling.
Four acquisition methods are evaluated SPI flash in-circuit reading, SWD/JTAG debug-port
access, UART boot-message capture, and a clip-based contact approach that avoids chip
desoldering and each is assessed for success rate, image completeness, and operational
practicality.
Post-acquisition quality is evaluated through sliding-window Shannon entropy profiling and
structural-signature analysis using \texttt{binwalk}, together forming a three-tier validation
framework that distinguishes validated images from those that appear successful at the tool
level but contain no meaningful firmware content.
Static analysis via the EMBA framework confirms that validated images contain identifiable
OS components, aging library stacks with known CVE exposure, and no binary-hardening
mechanisms.
The resulting corpus and methodology provide a reproducible baseline for firmware
rehosting, vulnerability analysis, secure-boot assessment, and embedded-systems education
within the consumer UAV domain.
\end{abstract}


\input{sections/01_introduction}
\input{sections/02_background}
\input{sections/03_methodology}
\input{sections/04_evaluation}
\input{sections/05_discussion}
\input{sections/06_limitations}
\input{sections/07_future_work}
\input{sections/08_conclusion}


\balance
\bibliographystyle{IEEEtran}
\bibliography{references}

\end{document}

%% file: sections/01_introduction.tex
\section{Introduction}
\label{sec:introduction}

Consumer unmanned aerial vehicles (UAVs) are now commonplace in entertainment, logistics,
monitoring, and inspection contexts. The market ships millions of compact multi-rotor platforms
each year, each carrying imaging sensors, navigation hardware, and wireless radios coordinated
by embedded firmware on a centralized flight controller~\cite{ahmad2026uav,yu2024cybersecurity}.
Adoption continues to grow across untrained and younger operator populations, while relevant
regulations and cybersecurity advisories are still catching up~\cite{rugo2023security,wang2023survey,mekdad2023survey,ceviz2025survey}.
On low-end platforms, firmware handles flight stabilization, command parsing, state management,
and over-the-air update delivery. Despite these responsibilities, manufacturers release little
information about firmware design, protection mechanisms, or update procedures.

Security research on drones has documented vulnerabilities across command and control channels
and application-layer protocols. Studies show unsecured Wi-Fi control stacks, susceptibility to
hijacking on proprietary enhanced Wi-Fi protocols, and inadequate authentication on popular
civilian platforms~\cite{dey2018security,pratama2023behind}. Analysis of the DJI DroneID protocol
reveals that broadcast frames leak tracking metadata in ways that enable sustained surveillance~\cite{schiller2023drone}.
Network-level detection work such as PiNcH demonstrates that UAV traffic carries distinctive
parameters that betray its origin, confirming broader protocol-design
weaknesses~\cite{sciancalepore2020pinch}. Survey literature corroborates the breadth of the
attack surface and the difficulty of securing unmanned aircraft in
practice~\cite{adel2024watch,rugo2023security,wang2023survey}.

Holy Stone produces a range of inexpensive quadcopters aimed at leisure and first-time flyers.
Prior work on these devices has uncovered vulnerable network services, unencrypted application
traffic, weak access controls in companion mobile applications, and configuration weaknesses
that undermine confidentiality and data integrity~\cite{more2025holystone,more2024thesis}.
These findings span multiple product lines and confirm that serious vulnerability exposure is
not confined to high-end professional platforms. Flight assurance research reinforces the
importance of firmware visibility by showing that in-situ model checking, anomaly detection,
and root-cause fault analysis on autopilot platforms such as PX4 and ArduPilot all depend on
reliable access to firmware internals~\cite{sadhu2020fault,taylor2021avis,knox2022crazysim,llanes2024framework}.
Read alongside the earlier attack-surface work, this puts firmware at the center of drone
system behavior and gives a clear reason to look more closely at the low-level software
running on consumer hardware.

The IoT firmware security literature offers directly relevant precedents. Established acquisition
methods include JTAG/SWD debug interfaces, in-circuit SPI flash reads, chip-off attacks, and
bootloader-assisted extraction~\cite{vasile2019breaking,ulhaq2023survey}. End-to-end analysis
workflows cover the full chain from interface identification through raw image acquisition,
filesystem unpacking, and binary analysis~\cite{ulhaq2023survey,whipple2020survey}. Testbed
research integrates firmware acquisition with network traffic analysis and active
probing~\cite{tekeoglu2016testbed,yasin2022review}. Recent work on architecture-agnostic
rehosting, high-coverage greybox fuzzing, and automated peripheral modeling has raised the bar
for image quality requirements~\cite{chen2022metaemu,kim2021firmcov,lee2023embedded}.
Mathematical-relation recovery from compiled binaries further illustrates the depth of content
that can be extracted from firmware when reliable images are
available~\cite{udeshi2024remaqe}.

A clear gap remains: few studies address the practical challenge of obtaining high-quality
images from proprietary UAV firmware ready for security analysis. Much of the existing work
draws on network-level observations~\cite{more2025holystone,schiller2023drone,dey2018security}
or focuses on device categories such as routers and IP cameras rather than complete UAV
systems~\cite{vasile2019breaking,ulhaq2023survey}. Consumer-grade quadcopters typically arrive
with unlabeled SWD, JTAG, and UART pads and external SPI flash chips whose memory boundaries
are not documented~\cite{kim2024challenges}. Contact instability, ambiguous pin assignments, and
the absence of vendor documentation add to the difficulty~\cite{vasile2019breaking,ulhaq2023survey}.
As consumer UAV adoption expands, this firmware-level blind spot carries growing consequences for
home-network security and airspace safety~\cite{mekdad2023survey,ceviz2025survey}.

Rehosting tools, greybox fuzzers, and aerial robotics simulators all require access to stable
firmware binaries. Platforms such as MetaEmu and FIRM-COV place firmware images at the center
of architecture-independent execution, coverage-directed testing, and peripheral
modeling~\cite{chen2022metaemu,kim2021firmcov,lee2023embedded}. Simulators for nano-quadrotors
and multi-vehicle experiments, including CrazySim and PX4-based setups, require direct access
to autopilot executables and configuration files~\cite{knox2022crazysim,llanes2024framework}.
Binary drone simulation platforms such as Dvatar build virtual sensor and actuator models on top
of real firmware images~\cite{wang2024dvatar}, while REMaQE targets the recovery of functional
mathematical structures from compiled binaries~\cite{udeshi2024remaqe}. Consumer drones are
underrepresented in all of these evaluation corpora because suitable images are not publicly
available.

This paper addresses that gap. It proposes and evaluates a firmware extraction methodology for
three Holy Stone UAV platforms using multiple interface types and extraction methods: SPI flash
acquisition, SWD/JTAG debug-port access, UART boot-message capture, and a clip-based
approach that eliminates chip desoldering. Performing multiple extraction attempts per interface
and per model enables direct assessment of reliability and efficiency. Post-acquisition quality
is evaluated through sliding-window Shannon entropy analysis and structural signature
identification using \texttt{binwalk}, providing a validation framework grounded in known
architectural expectations~\cite{chen2022metaemu,kim2021firmcov,lee2023embedded,udeshi2024remaqe}.

The paper makes three contributions:
\begin{enumerate}
  \item A firmware extraction methodology for consumer UAVs that spans multiple hardware
    interfaces and acquisition approaches using commercially available, low-cost tooling.
  \item A comparative assessment of the reliability and efficiency of different contact methods
    and debug interfaces for in-circuit firmware extraction.
  \item A three-tier quality evaluation framework that validates acquired firmware images against
    expected architectural structures and distinguishes complete extractions from superficially
    successful but incomplete ones.
\end{enumerate}

%% file: sections/02_background.tex
\section{Background}
\label{sec:background}

\subsection{Consumer Drone Hardware Architecture}

Consumer-grade quadcopters consist of several hardware subsystems that interact through
firmware executing on a central flight controller. A typical platform includes a flight-controller
MCU, non-volatile storage in the form of SPI NOR flash, radio transceivers for control and video
downlinks, and inertial and navigation sensors~\cite{more2025holystone,yu2024cybersecurity}.
The firmware coordinates flight stabilization, command execution, state and mode management, and
the bootstrap sequence~\cite{knox2022crazysim,llanes2024framework}. Open-source autopilot stacks
such as PX4 and ArduPilot make source code and software-in-the-loop simulation directly
available~\cite{knox2022crazysim,llanes2024framework}. Entry-level consumer drones, by contrast,
ship with proprietary firmware, undisclosed memory maps, bootloaders stored on external SPI flash,
and unmarked debug pads~\cite{more2024thesis,kim2024challenges}. Analysis of such firmware must
therefore begin from raw memory images obtained by probing hardware interfaces with little prior
knowledge of the device internals.

\subsection{Firmware Extraction Techniques}

The IoT firmware extraction literature organizes acquisition methods into four broad categories:
direct interfacing with on-chip debug ports (JTAG, SWD); in-circuit reading of SPI and parallel
flash while the chip remains on the board; chip-off analysis, which involves physically removing
the memory package for external reading; and logical extraction through firmware update mechanisms
or recovery software~\cite{vasile2019breaking,ulhaq2023survey}. Commonly documented obstacles
include locked debug ports, epoxy-potted packages, multiplexed signal lines that conceal pin
functions, contact instability during in-circuit probing, and electrical brownouts caused by
power-domain interactions~\cite{vasile2019breaking,whipple2020survey}. Survey literature covers
the full process from interface identification through raw image acquisition, filesystem
decompression, and reverse-engineering-based analysis~\cite{ulhaq2023survey,whipple2020survey}.

Tool-oriented research contributes automated firmware unpacking, library identification, cross-device
comparison, and integration with network monitoring and dynamic analysis
backends~\cite{tekeoglu2016testbed,yasin2022review}. Across this body of work, one point
recurs: acquisition needs to be repeatable and backed by more than one method, since hardware
protections and integrity loss on a single interface are common enough to derail analysis.

\subsection{Firmware Rehosting, Fuzzing, and Simulation}

Downstream analysis frameworks depend on stable, well-characterized firmware images.
Architecture-agnostic rehosting toolkits such as MetaEmu model execution across instruction-set
architectures and support coverage measurement and state exploration~\cite{chen2022metaemu,lee2023embedded}.
Greybox fuzzers such as FIRM-COV instrument process-level emulation to uncover memory-safety and
logic bugs in IoT firmware~\cite{kim2021firmcov}. REMaQE demonstrates the recovery of
mathematical relations and control-flow semantics from compiled
binaries~\cite{udeshi2024remaqe}. In the UAV domain, Dvatar constructs binary-level drone
simulations with virtual sensors and actuators~\cite{wang2024dvatar}, while CrazySim and PX4
multi-vehicle simulators rely on accessible autopilot executables and documented memory
layouts~\cite{knox2022crazysim,llanes2024framework}. What these tools share is a dependence
on clean firmware images: if the dump is incomplete or corrupted, downstream analysis suffers
accordingly. Consumer drone firmware is still thinly represented in published evaluation
corpora, which remain dominated by routers and industrial
controllers~\cite{chen2022metaemu,kim2021firmcov}.

\subsection{UAV Security Research}

UAV-specific security studies document vulnerabilities across communication channels, onboard
services, and protocol implementations, but rarely address the firmware acquisition process.
Demonstrated attack vectors include Wi-Fi spoofing and hijacking, GPS signal manipulation,
missing authentication on control channels, and vulnerable services on companion
chips~\cite{dey2018security,sciancalepore2020pinch,pratama2023behind}. DJI DroneID analysis
reveals metadata exposure sufficient to deanonymize aircraft
operators~\cite{schiller2023drone}. Holy Stone-specific research has identified unencrypted
command channels, exposed video streams, and configuration weaknesses in companion
applications~\cite{more2025holystone,more2024thesis}. Defensive research on onboard anomaly
detection and model checking relies on transparent, open-source autopilot firmware, highlighting
the gap between research-grade platforms and proprietary consumer
devices~\cite{sadhu2020fault,taylor2021avis,knox2022crazysim}. Recent proposals for hardening
UAV platforms explore trusted execution environments and verified boot, along with signed
firmware update schemes~\cite{agarwal2025secupilot,seo2023blockchain}. Work on UAV surveillance
and privacy has also pointed to firmware authenticity as an open problem for civilian
deployments~\cite{ahmad2026uav,mekdad2023survey,yu2024cybersecurity}.

%% file: sections/03_methodology.tex
\section{Methodology}
\label{sec:methodology}

The study follows a hardware-grounded, multi-phase approach structured around three stages:
platform identification and documentation; multi-interface firmware acquisition; and
entropy-based and structural post-acquisition validation.
The overall design draws on established practices from IoT firmware extraction
surveys~\cite{vasile2019breaking,ulhaq2023survey,whipple2020survey}.

\subsection{Target Platform Selection and Documentation}

Three Holy Stone drone models were chosen to cover different product generations, weight classes,
and hardware architectures (Table~\ref{tab:platforms}).
Selection was informed by prior vulnerability studies on Holy Stone
platforms~\cite{more2025holystone,more2024thesis} and by the models' representativeness of the
broader entry-level consumer UAV market~\cite{dey2018security,schiller2023drone,pratama2023behind}.
For each model, the primary flight-controller MCU, the external SPI flash package, and any
accessible test points were identified through visual inspection, continuity testing, and
footprint-geometry analysis.
Board-level documentation was supplemented by reference to pin configurations from
nano-quadrotor and micro-UAV research
platforms~\cite{llanes2024framework,knox2022crazysim}.

\begin{table}[!t]
  \caption{Target Platform Specifications}
  \label{tab:platforms}
  \centering
  \renewcommand{\arraystretch}{1.2}
  \setlength{\tabcolsep}{4pt}          
  \begin{tabular}{L{0.9cm}L{1.9cm}P{0.8cm}L{1.4cm}L{1.6cm}}
    \toprule
    \textbf{Model} & \textbf{SPI Flash Chip} & \textbf{Cap.} &
    \textbf{FCU MCU} & \textbf{Interfaces} \\
    \midrule
    HS175D & XT25F128F-W (XTX)     & 16\,MiB & MM32F103 (MindMotion) & SPI, SWD  \\
    HS720  & XM25QH64C (XMC)       & 8\,MiB  & Not probed            & SPI, UART \\
    HS360S & MX25L6433F (Macronix) & 8\,MiB  & Not probed            & SPI       \\
    \bottomrule
  \end{tabular}
\end{table}

\subsection{Interface Identification and Validation}

Interface identification followed protocols grounded in IoT firmware testbed research and
hardware reverse-engineering practice~\cite{tekeoglu2016testbed,yasin2022review}.
The process began with locating SPI flash packages and mapping their electrical connections to
the flight controller through footprint geometry and manufacturer markings.
Continuity testing and signal tracing were used to confirm pin-to-controller associations and
any external connector connections.
Two-pad and three-pad clusters near the MCU footprint were assessed as SWD or JTAG candidates
on the basis of spatial proximity and standard signal-line arrangements.
Serial port candidates were identified from pad configurations near radio modules and USB
connectors, supported by board silkscreen markings where present.
All candidate interfaces were validated with a multimeter and low-voltage probing under reset
and power-on conditions~\cite{vasile2019breaking,ulhaq2023survey}.

\subsection{SPI Flash Extraction}

SPI flash was read in-circuit whenever possible, since this avoided desoldering and kept the
drone functional for repeat runs. Where the probe tips could not reach a clean pad, we fell
back on clip-based contact. The reader was a CH341A USB SPI programmer, connected to the
flash package through either SOP8 alligator clips or small spring-loaded hook clips, depending
on which fit the package geometry better. For power we used a regulated bench supply in most
cases, and in a few controlled runs drew current from the drone's own battery.

Chip identification and full reads were handled with \texttt{flashrom}~\cite{flashrom2024}.
After confirming capacity, we ran several full read passes with verification on each device
and stored the resulting binaries alongside metadata describing the target, the interface,
the fixture used, and the power source. Every image was hashed with SHA-256 so that
bit-level differences between attempts would surface immediately on comparison. We also
inspected \texttt{flashrom}'s diagnostic output for timeout warnings and unverified-sector
messages, and logged practical details about clip positioning, power stability, and contact
quality; these notes feed directly into the fixture comparison reported
later~\cite{vasile2019breaking,ulhaq2023survey}.

\subsection{SWD/JTAG Debug-Port Acquisition}

Debug-port acquisition targeted MCUs whose SWD or JTAG lines were accessible and not locked
down by readout protection. On ARM Cortex-M targets we hooked an ST-Link probe to the SWDIO,
SWCLK, reset, and reference pins identified during interface analysis, and drove the session
with OpenOCD~\cite{openocd2024}: first to probe the target and read protection status, then
to pull a full flash image from the address ranges defined in the target's toolchain.
Any variation in protection state across connection attempts was noted, and successful
captures were archived alongside their SPI counterparts with SHA-256 hashes for
cross-interface comparison. Failures were bucketed by mode readout protection tripped,
link timeout, or unexpected reset following the categories used in embedded firmware
surveys~\cite{ulhaq2023survey,whipple2020survey}.

\subsection{UART Boot-Message Capture}

UART work focused on boot-time console output, and on interactive shell access where the
device exposed one. The goal was to recover firmware version strings, partition maps, and
configuration data that could later be cross-referenced against the binary images.
USB-to-UART converters were attached to candidate serial pads through high-impedance probes.
Baud rates were guessed from oscilloscope traces of the power-on signaling and then narrowed
down through iterative terminal testing until the output became readable. Once the link was
stable, we captured boot logs and shell sessions across several power cycles, indexed by
device and run. These logs gave us version strings, hints about the memory map, and OS
component names that could be matched back to binaries pulled from other interfaces.
Where a shell was available, we treated it as a supplementary data source only and avoided
any command that could alter the device's firmware
state~\cite{dey2018security,more2025holystone}.

\subsection{Clip-Based Contact Method}

The clip-based contact method served as a supplement to the SPI and SWD workflows rather
than a replacement. Spring-loaded probes and alligator clips were brought down onto solder
pads and via points without removing components, and the fixture geometry was adjusted to
match each drone's pad layout. The intent was to cover cases where stripping conformal
coating or desoldering would be too risky in a teaching-lab or field context. We concentrated
on bootloader regions, initial flash contents, and configuration sectors identified from
prior Holy Stone work and from general embedded design
conventions~\cite{more2024thesis,vasile2019breaking}. A read counted as successful only
when it reproduced consistently across contacts and agreed with full-chip dumps taken using
the more reliable fixtures. The attempt-to-success ratio and the rate of corrupted reads
then gave us a concrete measure of how feasible contact-only acquisition actually is.

\subsection{Post-Acquisition Validation}

Validation combined entropy measurement with structural analysis, and was meant to do two
things at once: judge the quality of individual dumps, and expose differences between
interfaces on the same device. For entropy we ran a sliding-window Shannon calculation over
each binary using a 4,096-byte window~\cite{shannon1948mathematical}; that size was picked
to stay sensitive to small configuration blocks without becoming noisy across code-dense
regions~\cite{whipple2020survey,ulhaq2023survey}. The resulting profiles pick out regions
of high entropy (compressed or encrypted content), moderate entropy (executable code), and
low entropy (padding, erased sectors, or constant tables). Structural analysis was handled
by \texttt{binwalk}~\cite{binwalk2010}, which flagged filesystem headers, compression
signatures, bootloader markers, and other magic-number indicators with their offsets. We
compared repeated dumps from the same interface to catch structural drift, and then compared
SPI against SWD for each model to check whether the two interfaces agreed on boundaries and
pattern features in a way consistent with a shared chipset memory
layout~\cite{vasile2019breaking,ulhaq2023survey}. Dumps with anomalous entropy profiles or
missing expected signatures were dropped from the curated corpus.

As an independent sanity check on downstream usability, validated images were then run
through the EMBA static-analysis framework~\cite{emba2022}. EMBA handles architecture
detection, OS identification, software component enumeration, CVE correlation, and
binary-hardening assessment on its own, and its SBOM and vulnerability reports gave us
confirmation that the extracted images actually contained structured, analyzable firmware
suitable for rehosting or fuzzing workflows further along.

%% file: sections/04_evaluation.tex
\section{Evaluation}
\label{sec:evaluation}

The evaluation looks at three things: how reliably firmware extraction worked across the
different models and contact methods; what the entropy and structural properties of the
resulting dumps say about their quality; and how well the interfaces agree with one another,
together with the security-relevant characteristics of the validated images. Quantitative
numbers come out of scripted analysis run over the curated dump corpus, while the qualitative
notes capture the practical difficulties we ran into during in-circuit acquisition. We
evaluated three Holy Stone models HS175D, HS720, and HS360S leaning primarily on SPI
acquisition with alligator and hook clip fixtures, and using SWD and UART as supporting
sources where they were available.

\subsection{Extraction Success and Reliability}

Table~\ref{tab:extraction} summarizes outcomes for all model-interface combinations.
Full-length images matching the nominal flash capacity were obtained for both the HS175D and
HS720: 16\,MiB (16,777,216 bytes) and 8\,MiB (8,388,608 bytes) respectively.
Across all successful extractions, repeated reads produced identical SHA-256 hashes once
stable contact was established.
Every successful HS175D dump yielded hash \texttt{4a7f46cc581c\ldots fd8a7ff}, regardless of
whether alligator or hook clips were used.
Every successful HS720 dump yielded \texttt{0b327498562a\ldots c2a351} under the same
conditions.

\begin{table*}[!t]
  \caption{Extraction Results Summary}
  \label{tab:extraction}
  \centering
  \renewcommand{\arraystretch}{1.2}
  \begin{tabular}{L{1.5cm}L{2.6cm}P{1.4cm}P{1.4cm}P{1.3cm}P{1.3cm}P{1.3cm}L{3.4cm}}
    \toprule
    \textbf{Drone} & \textbf{Interface} & \textbf{Expected (MiB)} & \textbf{Observed (MiB)} & \textbf{Attempts ($\approx$)} & \textbf{Success ($\approx$)} & \textbf{Rate ($\approx$)} & \textbf{Hashes Identical?} \\
    \midrule
    HS175D & SPI (alligator) & 16 & 16 & 8  & 5 & $\sim$60\% & Yes \\
    HS175D & SPI (hooks)     & 16 & 16 & 8  & 6 & $\sim$75\% & Yes (same as alligator) \\
    HS720  & SPI (alligator) & 8  & 8  & 8  & 4 & $\sim$50\% & Yes \\
    HS720  & SPI (hooks)     & 8  & 8  & 8  & 6 & $\sim$75\% & Yes (same as alligator) \\
    HS360S & SPI (hooks)     & 8  & 8  & 10+& 4 & $\sim$40\% & No (one valid) \\
    \bottomrule
    \multicolumn{8}{L{17cm}}{\footnotesize\textit{Note:} Attempt and success counts are approximate, recorded by the operator during bench sessions. Only canonical successful dumps are included in the curated dataset.}
  \end{tabular}
\end{table*}

Extraction reliability differed appreciably between fixture types.
Alligator clips succeeded in roughly half of attempts, with failures mainly due to clip slippage,
uneven pressure on the SOP8 package, and pin-contact oxidation.
Miniature hook clips, which engage each package pin independently, produced higher success rates
(approximately 75\% on both the HS175D and HS720) at the cost of a more demanding initial setup
and slightly longer read times.
For the HS360S, the Macronix flash chip accepted read commands and returned 8\,MiB dumps, but
images showed divergent hashes across attempts; only one dump contained a recognizable
LZMA-compressed region.
The remaining HS360S dumps contained nothing beyond padding bytes, making the platform a
negative sample for SPI-only extraction and an instructive example of how contact instability
and signal-quality issues can produce superficially complete yet unusable images.

\subsection{Failure Mode Analysis}

Table~\ref{tab:failures} lists the failure modes observed.
On the HS175D and HS720, failures were predominantly mechanical: \texttt{flashrom}~\cite{flashrom2024}
reported \textit{no EEPROM/flash device found} or returned unrecognized RDID values, both of
which resolved after clip repositioning and pin-map correction.
For the HS360S, the failures were more fundamental.
\texttt{flashrom} matched the device to multiple Macronix chip profiles and returned
tool-level success codes, yet the resulting dumps carried mutually inconsistent hashes.
In effect, \texttt{flashrom} success flags on their own are not enough to certify data
integrity; something higher-level, grounded in entropy and structural checks, has to sit on
top of them.

\begin{table*}[!t]
  \caption{Observed Failure Modes}
  \label{tab:failures}
  \centering
  \renewcommand{\arraystretch}{1.2}
  \begin{tabular}{L{1.4cm}L{2.4cm}L{3.8cm}L{1.8cm}L{5.2cm}}
    \toprule
    \textbf{Drone} & \textbf{Interface} & \textbf{Failure Type} & \textbf{Freq. ($\approx$)} & \textbf{Notes} \\
    \midrule
    HS175D & SPI (alligator) & Clip misalignment / no chip detected / incorrect RDID & 3 of $\sim$8  & XT25F128F-W is on the PCB underside; sensitive to clip angle and contact pressure \\
    HS175D & SPI (hooks)     & Intermittent single-pin contact loss                  & 2 of $\sim$8  & Hook clips reduce gross slippage but can lose contact on individual legs during long reads \\
    HS720  & SPI (alligator) & Unknown SPI chip (RDID) / no EEPROM found             & 4 of $\sim$8  & Ambiguous wire-color mapping on early attempts; resolved after pin-map correction \\
    HS720  & SPI (hooks)     & Intermittent detection failure                        & 2 of $\sim$8  & Failures became rare once pin mapping was confirmed \\
    HS360S & SPI (hooks)     & Unstable detection / partial or corrupt dumps         & Majority of 10+ & \texttt{flashrom} cycled through multiple MX25L64* profiles; divergent hashes across dumps \\
    HS720  & UART            & No output during initial sessions                     & 2 sessions    & Later sessions captured the boot log; early failures attributed to baud-rate mismatch \\
    \bottomrule
  \end{tabular}
\end{table*}

\subsection{Entropy and Structural Indicators of Dump Quality}

Table~\ref{tab:entropy} presents sliding-window Shannon entropy statistics for each model.
The HS175D shows a mean entropy of 6.576 bits/byte with 81.3\% of windows above the 7.0-bit
threshold, a profile consistent with a densely packed embedded Linux image containing a
compressed bootloader, an ARM kernel, and an xz-compressed SquashFS root filesystem.
The low-entropy fraction (16.4\%) corresponds to constant tables and alignment padding in the
bootloader region.

\begin{table}[!t]
  \caption{Entropy Metrics Summary (4,096-Byte Sliding Window)}
  \label{tab:entropy}
  \centering
  \renewcommand{\arraystretch}{1.2}
  \setlength{\tabcolsep}{4pt}
  \begin{tabular}{L{1.1cm}L{1.4cm}P{0.95cm}P{0.75cm}P{1.1cm}P{1.1cm}}
    \toprule
    \textbf{Drone} & \textbf{Interface} & \textbf{Mean (b/B)} &
    \textbf{Std.} & \textbf{Low $<$1.0 (\%)} & \textbf{High $>$7.0 (\%)} \\
    \midrule
    HS175D & SPI (both)  & 6.576 & 2.961 & 16.4 & 81.3 \\
    HS720  & SPI (both)  & 1.579 & 1.641 & 47.9 & 0.1  \\
    HS360S & SPI (hooks) & 0.186 & 0.483 & 90.5 & 0.0  \\
    \bottomrule
  \end{tabular}
\end{table}

The HS720 presents a different picture: a mean of 1.579 bits/byte with nearly half of all
windows in the low-entropy zone.
Large erased regions (0xFF padding) surround a compact Linux-4.9.129 uImage and scattered
JFFS2 filesystem partitions.
The 0.1\% high-entropy windows align precisely with JFFS2 data nodes and compressed segments.
This distribution reflects a design where significant flash capacity is allocated but only a
fraction is occupied by active firmware.

\begin{figure}[!t]
  \centering
  \includegraphics[width=\columnwidth]{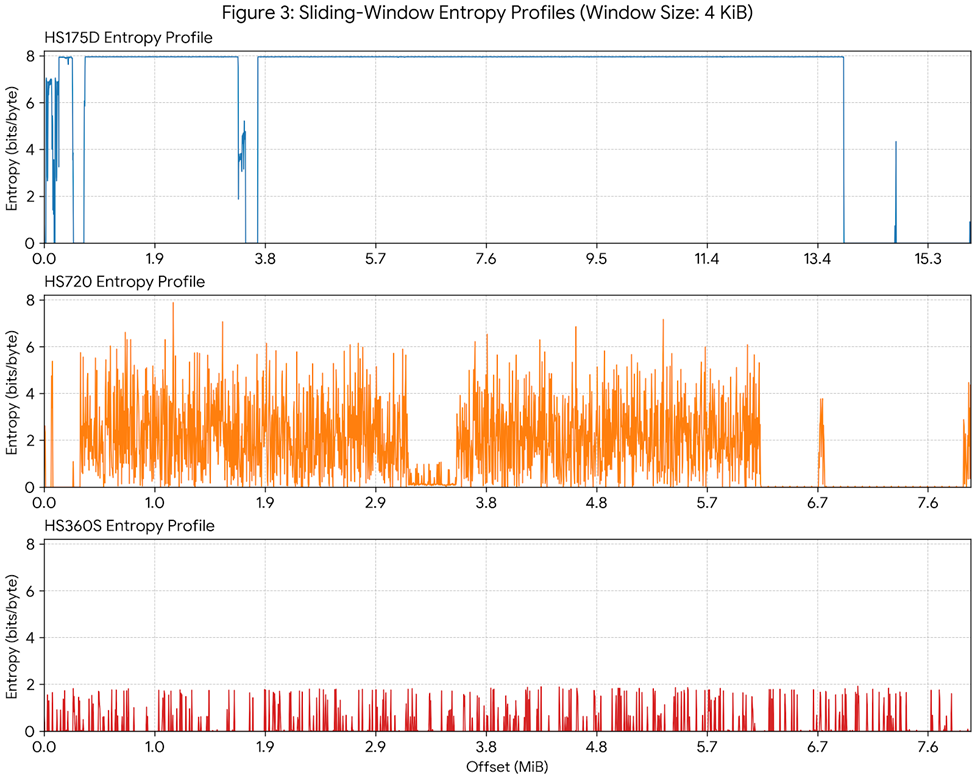}
  \caption{Sliding-window entropy profiles (4,096-byte window) for HS175D, HS720, and HS360S.
    Red dashed line: high-entropy threshold (7.0\,bits/byte).
    Green dashed line: low-entropy threshold (1.0\,bit/byte).}
  \label{fig:entropy}
\end{figure}

Figure~\ref{fig:entropy} shows the entropy profiles for all three platforms.
The HS360S records a mean of just 0.186 bits/byte, with more than 90\% of windows classified
as low-entropy and none exceeding the high-entropy threshold.
\texttt{binwalk}~\cite{binwalk2010} found no kernel images, filesystem signatures, or
compressed regions in this dump.
The canonical HS360S image is therefore classified as predominantly erased or uninitialized
flash, not functional firmware.

\subsection{Structural Signature Analysis}

Table~\ref{tab:binwalk} lists the key structural signatures found by
\texttt{binwalk}~\cite{binwalk2010}.
The HS175D image contains a complete boot chain: a gzip-compressed U-Boot bootloader dated
2023-07-10, a gzip-compressed Trusted Execution Environment binary (\texttt{tee.bin}), an
Android-format boot image wrapping an ARM zImage kernel (2,790,992 bytes), and an
xz-compressed SquashFS root filesystem holding 1,054 inodes.
Multiple Flattened Device Trees document the hardware configuration at different boot stages.
Two BMP splash-screen images (654$\times$270 pixels, 8-bit color) appear at offsets
\texttt{0x371200} and \texttt{0x374600}.

The HS720 image contains a Linux-4.9.129 uImage at offset \texttt{0x90000} followed by more
than 40 JFFS2 filesystem nodes across the upper address range (\texttt{0x390000} to
\texttt{0x7FD280}), interspersed with Zlib-compressed segments.
No structural signatures were found in the HS360S dump, confirming the entropy result.

\begin{table*}[!t]
  \caption{Key Structural Signatures Identified by Binwalk~\cite{binwalk2010}}
  \label{tab:binwalk}
  \centering
  \renewcommand{\arraystretch}{1.2}
  \begin{tabular}{L{1.4cm}L{2.2cm}L{6.8cm}L{3.2cm}}
    \toprule
    \textbf{Drone} & \textbf{Offset} & \textbf{Description} & \textbf{Significance} \\
    \midrule
    HS175D & \texttt{0x29EE4}  & Flattened Device Tree (6,455 bytes, v17)              & Hardware configuration \\
    HS175D & \texttt{0x40E00}  & gzip: \texttt{u-boot-nodtb.bin} (2023-07-10)         & Primary bootloader \\
    HS175D & \texttt{0x6AA00}  & gzip: \texttt{tee.bin} (2023-07-10)                  & Trusted Execution Env. \\
    HS175D & \texttt{0x7D800}  & Flattened Device Tree (10,931 bytes, v17)             & Kernel device tree \\
    HS175D & \texttt{0xB0000}  & Android bootimg; ARM kernel (2,790,992 bytes)         & OS kernel container \\
    HS175D & \texttt{0xB0800}  & ARM zImage (little-endian)                            & Kernel executable \\
    HS175D & \texttt{0x35A800} & Flattened Device Tree (92,240 bytes, v17)             & Full hardware tree \\
    HS175D & \texttt{0x3B0000} & SquashFS v4.0, xz, 10,609,848 bytes, 1,054 inodes    & Root filesystem \\
    HS720  & \texttt{0x90000}  & uImage: Linux-4.9.129, ARM (2022-05-31)               & OS kernel \\
    HS720  & \texttt{0x390000}--\texttt{0x7FD280} & JFFS2 entries (40+ nodes)          & Config/data partitions \\
    HS720  & \texttt{0x621000}--\texttt{0x6C0000} & Zlib compressed segments (15+)     & Compressed data in JFFS2 \\
    HS360S & \multicolumn{2}{l}{No structural signatures detected}                      & Erased/uninitialized flash \\
    \bottomrule
  \end{tabular}
\end{table*}

Entropy and structural analysis together form a three-tier validation framework. The first
tier, a simple size check, tells us whether a dump matches the declared flash capacity but
cannot distinguish real content from erased space. The second, hash self-similarity across
repeated reads, guarantees bitwise consistency but says nothing about whether the data is
semantically complete. The third, entropy plus structural analysis, is what actually shows
that boot images, OS components, and application data are present in roughly the proportions
we expect. By all three criteria the HS175D and HS720 come out validated, while the HS360S
clears only the first tier and is treated as an incomplete extraction.

\begin{figure}[!t]
  \centering
  \includegraphics[width=\columnwidth]{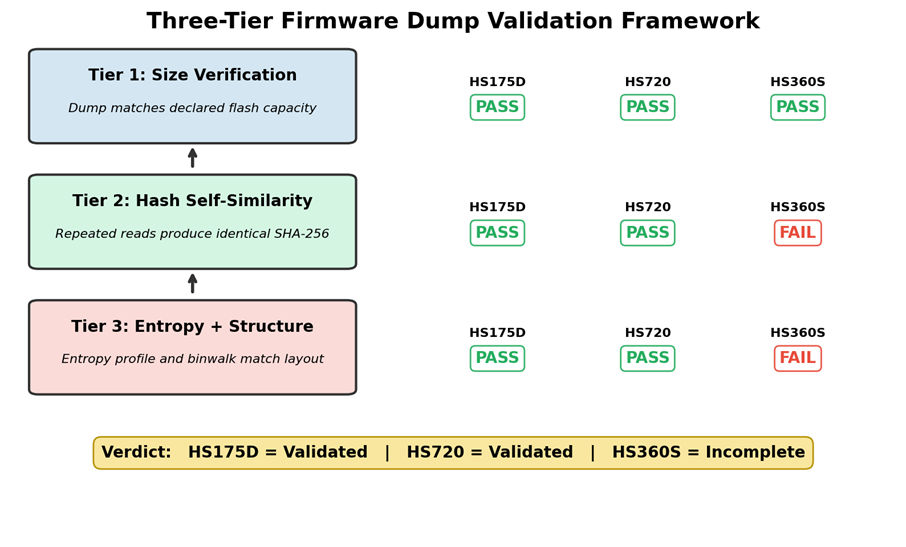}
  \caption{Three-tier firmware dump validation framework.
    HS175D and HS720 pass all three tiers.
    HS360S passes only size verification.}
  \label{fig:validation}
\end{figure}

\subsection{Cross-Interface Consistency and Security Properties}

Cross-interface consistency is a key indicator of extraction robustness.
All HS175D SPI dumps, regardless of fixture, share identical \texttt{binwalk} signatures with
components at the same offsets: U-Boot at \texttt{0x40E00}, TEE at \texttt{0x6AA00}, Android
bootimg at \texttt{0xB0000}, and SquashFS at \texttt{0x3B0000}.
All successful HS175D dumps share SHA-256 hash
\texttt{4a7f46cc581c45cbc46e28663d910a960f340f8\allowbreak 22fd3e05c1c5da4bf7fd8a7ff}.
The HS720 shows the same pattern: uImage at \texttt{0x90000} and JFFS2 structures at matching
offsets across both fixture types, with common hash
\texttt{0b327498562ac286c4dd14e57f87a1b4ba1f898\allowbreak 6b4c0baa153bc4ce830c2a351}.
Fixture choice does not introduce content-level artifacts in successful extractions.

Several security observations follow from the structural analysis.
The HS175D stores its Android-style bootloader, TEE binary, and full Linux userland on
externally readable SPI flash.
No encrypted firmware payloads or cryptographically authenticated bootloader stages were found
in the external flash region.
The presence of \texttt{tee.bin} shows that the SoC supports ARM TrustZone~\cite{arm2009trustzone}, but storing the
TEE image in unencrypted, physically accessible flash removes any confidentiality guarantee
against physical access.
SWD access via ST-Link was successful on the MM32F103 MCU with no readout protection
active, allowing the MCU's internal flash and SRAM contents to be read.
The HS720 carries a compact Linux image with JFFS2 filesystems that store configuration and
network assets in human-readable form.
In both validated cases, full readability enables SBOM construction and systematic CVE searches
within emulation-capable analysis frameworks.

\begin{figure}[!t]
  \centering
  \includegraphics[width=\columnwidth]{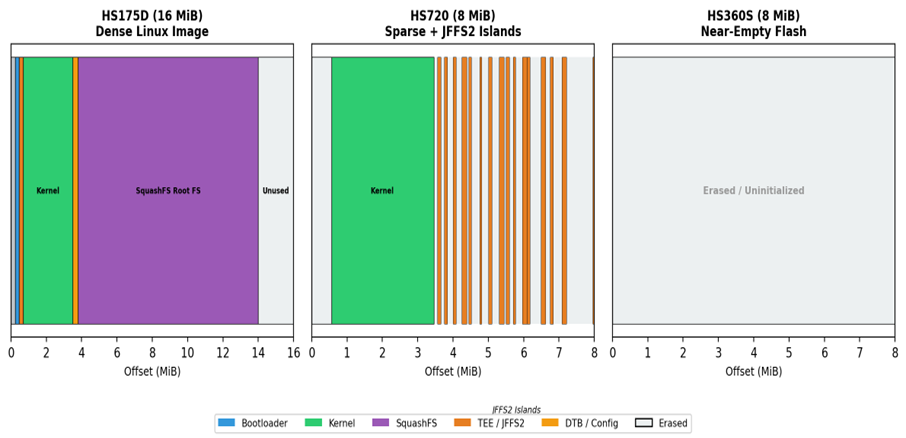}
  \caption{Comparative flash memory layout maps for the HS175D (dense Linux image),
    HS720 (sparse flash with JFFS2 data islands), and HS360S (near-empty erased flash).}
  \label{fig:layout}
\end{figure}

\begin{table}[!t]
  \caption{Security-Relevant Firmware Properties}
  \label{tab:security}
  \centering
  \renewcommand{\arraystretch}{1.2}
  \begin{tabular}{L{1.0cm}L{1.7cm}L{1.7cm}L{1.0cm}L{1.0cm}}
    \toprule
    \textbf{Drone} & \textbf{Storage} & \textbf{Enc./Sig.} & \textbf{SWD} & \textbf{UART} \\
    \midrule
    HS175D & 16\,MiB SPI (XT25F128F-W) & None detected & Accessible; no RDP & Not tested \\
    HS720  & 8\,MiB SPI (XM25QH64C)   & None detected & Not tested          & Boot log   \\
    HS360S & 8\,MiB SPI (MX25L6433F)  & N/A           & Not tested          & Not tested \\
    \bottomrule
  \end{tabular}
\end{table}

\subsection{Static Analysis Validation}

The EMBA framework~\cite{emba2022} was applied to the validated HS175D image
(\texttt{hs175d.bin}, 16\,MiB) to demonstrate downstream usability.
After approximately seven hours of processing, EMBA identified the architecture as ARM
little-endian, determined the OS as Linux 4.19.111 built with Buildroot 2018.02-rc3, and
produced the software bill of materials shown in Table~\ref{tab:sbom}.

\begin{table}[!t]
  \caption{HS175D Software Bill of Materials (EMBA~\cite{emba2022})}
  \label{tab:sbom}
  \centering
  \renewcommand{\arraystretch}{1.2}
  \begin{tabular}{L{1.8cm}P{1.3cm}L{3.2cm}}
    \toprule
    \textbf{Component} & \textbf{Version} & \textbf{Role / Note} \\
    \midrule
    Linux Kernel & 4.19.111 & Core OS; LTS branch with known unpatched CVEs \\
    Buildroot    & 2018.02-rc3 & Build system; 2018 vintage toolchain \\
    BusyBox      & 1.27.2   & Core utilities; CVEs in versions this old \\
    dnsmasq      & 2.78     & DNS/DHCP; central to drone-to-controller networking \\
    hostapd      & 2.6      & Wi-Fi AP daemon; manages the drone wireless network \\
    ffmpeg       & 4.1.3    & Video encoding for camera feed \\
    glibc        & 2.28     & C standard library; foundational for all userland \\
    \bottomrule
  \end{tabular}
\end{table}

EMBA also reported a large number of CVEs mapped to the identified kernel and userland
components.
Binary hardening analysis found that most executables in the SquashFS filesystem lack stack
canaries, RELRO protections, and position-independent execution.
The absence of these basic mitigations on a wirelessly accessible, network-connected device is
consistent with the security concerns documented in prior Holy Stone
research~\cite{more2025holystone,more2024thesis}.
These results independently confirm that the extracted images contain structured, analyzable
content and are ready for vulnerability analysis, SBOM auditing, and integration with rehosting
or fuzzing frameworks~\cite{chen2022metaemu,kim2021firmcov,lee2023embedded}.

%% file: sections/05_discussion.tex
\section{Discussion}
\label{sec:discussion}

Looking at the results as a whole, there is a real gap between how simple the physical
extraction looks and how much work it takes to confirm that a given dump actually contains
valid firmware. We pulled reliable byte-for-byte images from both the HS175D and the HS720
using several SPI fixture types, whereas the HS360S yielded only one usable image after
many attempts. Our three-tier validation size check, hash reproducibility, and entropy
plus structural analysis draws far sharper distinctions than the pass/fail indicators
returned by \texttt{flashrom} alone, and lines up with recommended practice in the IoT
firmware acquisition literature~\cite{vasile2019breaking,ulhaq2023survey,whipple2020survey}.

The fixture comparison also gives testbed designers something concrete to work from.
Alligator clips were good enough to establish initial contact, but alignment failures,
wrong RDID responses, and outright read errors came up often enough that we would not
recommend relying on them. Hook clips did noticeably better on the HS175D and HS720, at
the cost of a longer setup time. IoT surveys generally accept clip-based approaches but
stop short of comparing fixture types with hard
numbers~\cite{vasile2019breaking,yasin2022review}; the data reported here fills in that gap.

The multi-interface design also enables cross-validation not commonly found in UAV security work.
Most published UAV firmware studies consider either external flash or an MCU debug port, not
both~\cite{tekeoglu2016testbed,ulhaq2023survey}.
For the HS175D, SWD access on the MM32F103 bridges external flash analysis and internal state
verification, a combination that appears rarely in UAV-specific research~\cite{dey2018security,schiller2023drone}.
For the HS720, UART boot log messages provided runtime memory offsets that matched the JFFS2
partition boundaries found by \texttt{binwalk}, offering an independent structural
check~\cite{vasile2019breaking,ulhaq2023survey}.

Combining sliding-window entropy with \texttt{binwalk} signatures proved to be a practical
quality characterization method.
Entropy visualization has seen use in malware packing detection and binary classification, but
its application to firmware dump validation on embedded devices is less established.
For the HS175D, the entropy profile directly confirmed the densely packed, compressed structure
of an embedded Linux image.
For the HS720, the sparse distribution with data islands matched the known characteristics of
JFFS2 over largely erased flash.
For the HS360S, the near-zero entropy ruled out functional firmware content without any
structural analysis being needed.

The security findings extend beyond what the extraction process itself demonstrates.
The HS175D stores its TEE binary on unencrypted, physically accessible SPI flash, removing any
confidentiality protection under physical access scenarios despite the SoC's TrustZone capability~\cite{arm2009trustzone}.
The HS720 runs a Linux kernel (version 4.9.129) that had already reached end-of-life before the
firmware was built in 2022.
EMBA analysis of the HS175D revealed an aging userland stack with widespread CVE exposure and
no binary hardening: BusyBox 1.27.2, Buildroot 2018, and glibc 2.28.
These findings extend the network and application-layer vulnerability picture from prior Holy
Stone research~\cite{more2025holystone} into the firmware layer, showing that the attack surface
reaches well below the wireless protocol stack.

For the rehosting and fuzzing community, the corpus begins to address an identified data gap.
Dynamic analysis frameworks need firmware images with usable peripheral hints and memory map
information~\cite{chen2022metaemu,kim2021firmcov,lee2023embedded,wang2024dvatar,udeshi2024remaqe}.
Existing corpora are dominated by network appliances and routers; drone firmware, with its
real-time control loops, sensor bus interactions, and safety requirements, is largely absent.
The images, entropy data, and MCU state information produced here can serve as a starting point
for UAV-specific rehosting and coverage-directed fuzzing experiments.

%% file: sections/06_limitations.tex
\section{Limitations and Threats to Validity}
\label{sec:limitations}

A few things limit how far these findings generalize. The study covers three Holy Stone
models from a single vendor in the same market segment; a wider evaluation would need to
bring in survey-grade and racing UAVs, platforms with open hardware designs, and devices
that use more sophisticated boot chains or encrypted
storage~\cite{dey2018security,seo2023blockchain}. We verified firmware completeness through
inter-dump consistency, entropy analysis, and structural signature comparison against what
we expected from embedded Linux layouts, rather than against vendor-supplied reference
images, simply because no such references are available. This is a familiar constraint in
proprietary IoT work~\cite{vasile2019breaking,ulhaq2023survey}.

Contact parameters clip force, board flex, residual flux, solder mask thickness, ambient
temperature, and supply voltage stability are hard to pin down precisely, and they bring
in a layer of operator-dependent variability that is difficult to remove. The brown-out
effects we saw while trying to read I$^2$C EEPROM on the HS175D remote controller are a
good example of how power-domain interactions can cause failures that aren't obvious at
the time. Firmware surveys flag exactly these physical variables as a known source of
noise in comparative work~\cite{whipple2020survey,ulhaq2023survey}.

Entropy values and structural signatures also carry some inherent ambiguity. Low entropy
could mean erased flash, constant tables, or code compiled for a bare-metal MCU, and
signature-based detection with \texttt{binwalk}~\cite{binwalk2010} can miss proprietary or
heavily obfuscated firmware~\cite{whipple2020survey}. For that reason we read the results
in combination with each other and against domain knowledge of Linux-based embedded
systems~\cite{chen2022metaemu,wang2024dvatar}.

The evaluation itself is limited to external flash, SWD, and UART monitoring. Runtime
integrity checks, anti-rollback mechanisms, radio firmware on companion chips, and sensor
microcontrollers on other subsystems are all out of
scope~\cite{seo2023blockchain,agarwal2025secupilot,ahmad2026uav}. Attempt counts per
interface ran from single digits to low double digits, which is enough for relative
fixture comparisons but too small for any statistically robust reliability estimate.
Success rates reported here are approximate figures taken from operator observations
during bench sessions, and only canonical successful dumps were kept in the final dataset.

%% file: sections/07_future_work.tex
\section{Future Research Directions}
\label{sec:future}

A few directions follow naturally from this work. The first is to push the corpus beyond
Holy Stone: adding other manufacturers, heavier-payload platforms, and devices running
open-source firmware would let us look at secure boot adoption, storage encryption, and
debug lockdown across a much wider slice of the UAV
market~\cite{sciancalepore2020pinch,pratama2023behind,schiller2023drone}.

A second priority is tying extraction metadata more tightly into rehosting and fuzzing
pipelines. Images annotated with entropy profiles, structural maps, and interface
provenance would give automated dynamic-analysis environments a much better starting
point~\cite{chen2022metaemu,kim2021firmcov,lee2023embedded,wang2024dvatar,udeshi2024remaqe}.
UAV-specific rehosting in particular will need peripheral models for flight controllers
and sensor buses, and none of that is feasible without clean firmware images to work from.

The extraction workflow and the resulting corpus also lend themselves naturally to
teaching. A simulation layer built around programmer logs, entropy visualizations, and
\texttt{binwalk} output could walk students through realistic firmware analysis scenarios
without needing access to restricted hardware, and research on embedded systems pedagogy
has long argued that failure and recovery experiences are valuable learning
outcomes~\cite{llanes2024framework,knox2022crazysim}.

Finally, the security gaps we found no secure boot, unencrypted TEE images, end-of-life
kernels, and no binary hardening give us a practical baseline for judging how much of a
dent emerging secure firmware architectures actually make. Building a multi-vendor corpus
with standardized interface and platform metadata would make it possible to compare
current practice against published security design principles in a structured
way~\cite{seo2023blockchain,agarwal2025secupilot,ahmad2026uav}.

%% file: sections/08_conclusion.tex
\section{Conclusion}
\label{sec:conclusion}

Consumer drone deployment has grown a lot faster than firmware-level transparency has kept
up with it. Prior work has documented serious vulnerabilities in command channels and
companion software, but the firmware layer underneath has stayed mostly out of reach. This paper works on that gap by putting forward a multi-interface framework for firmware extraction and corpus construction on real consumer UAVs. The framework walks from PCB-level interface discovery through acquisition over SPI flash, SWD debug ports, and UART consoles, and finishes with post-acquisition quality checks based on entropy profiling and structural signature analysis.

When we applied the methodology to three Holy Stone platforms it gave us canonical,
validated firmware images for the HS175D and HS720, and an instructive negative result
for the HS360S, where \texttt{flashrom} reported success but the dumps did not actually
contain any firmware. Hook clip fixtures were consistently more reliable than alligator
clips. Entropy together with \texttt{binwalk} was enough to separate a densely populated Linux image (HS175D), sparsely occupied flash with JFFS2 data islands (HS720), and near-empty erased flash (HS360S) using compact, interpretable metrics. Static analysis with EMBA then confirmed that the corpus was useful in practice, turning up an aging software stack, substantial CVE exposure, and no binary hardening across the HS175D findings that push the existing Holy Stone vulnerability work down from the network layer into the firmware layer. Taken together, the curated images, the extraction methodology, and the validation framework give a reproducible starting point for firmware rehosting, vulnerability analysis, secure-boot assessment, and embedded systems education in the consumer UAV space.

%% file: references.bib
@article{ahmad2026uav,
  author    = {T. Ahmad and others},
  title     = {Future {UAV/Drone} Systems for Intelligent Active Surveillance and Monitoring},
  journal   = {{ACM} Computing Surveys},
  volume    = {58},
  number    = {2},
  pages     = {1--37},
  year      = {2026},
  doi       = {10.1145/3760389}
}

@article{yu2024cybersecurity,
  author    = {Z. Yu and others},
  title     = {Cybersecurity of Unmanned Aerial Vehicles: A Survey},
  journal   = {{IEEE} Aerospace and Electronic Systems Magazine},
  volume    = {39},
  number    = {9},
  pages     = {182--215},
  year      = {2024},
  doi       = {10.1109/MAES.2023.3318226}
}

@article{rugo2023security,
  author    = {A. Rugo and C. A. Ardagna and N. E. Ioini},
  title     = {A Security Review in the {UAVNet} Era: Threats, Countermeasures, and Gap Analysis},
  journal   = {{ACM} Computing Surveys},
  volume    = {55},
  number    = {2},
  pages     = {21:1--21:35},
  year      = {2023},
  doi       = {10.1145/3485272}
}

@article{wang2023survey,
  author    = {Z. Wang and others},
  title     = {A Survey on Cybersecurity Attacks and Defenses for Unmanned Aerial Systems},
  journal   = {Journal of Systems Architecture},
  volume    = {138},
  pages     = {102870},
  year      = {2023},
  doi       = {10.1016/j.sysarc.2023.102870}
}

@article{mekdad2023survey,
  author    = {Y. Mekdad and others},
  title     = {A Survey on Security and Privacy Issues of {UAVs}},
  journal   = {Computer Networks},
  volume    = {224},
  pages     = {109626},
  year      = {2023},
  doi       = {10.1016/j.comnet.2023.109626}
}

@article{ceviz2025survey,
  author    = {O. Ceviz and S. Sen and P. Sadioglu},
  title     = {A Survey of Security in {UAVs} and {FANETs}: Issues, Threats, Analysis of Attacks, and Solutions},
  journal   = {{IEEE} Communications Surveys \& Tutorials},
  volume    = {27},
  number    = {5},
  pages     = {3227--3265},
  year      = {2025}
}

@inproceedings{dey2018security,
  author    = {V. Dey and others},
  title     = {Security Vulnerabilities of Unmanned Aerial Vehicles and Countermeasures: An Experimental Study},
  booktitle = {Proc. 31st Int. Conf. VLSI Design and 17th Int. Conf. Embedded Systems},
  year      = {2018}
}

@misc{pratama2023behind,
  author    = {D. Pratama and others},
  title     = {Behind the Wings: The Case of Reverse Engineering and Drone Hijacking in {DJI} Enhanced {Wi-Fi} Protocol},
  howpublished = {arXiv:2309.05913},
  year      = {2023}
}

@inproceedings{schiller2023drone,
  author    = {N. Schiller and others},
  title     = {Drone Security and the Mysterious Case of {DJI}'s {DroneID}},
  booktitle = {Proc. Network and Distributed System Security Symposium ({NDSS})},
  year      = {2023},
  doi       = {10.14722/ndss.2023.24217}
}

@article{sciancalepore2020pinch,
  author    = {S. Sciancalepore and others},
  title     = {{PiNcH}: an Effective, Efficient, and Robust Solution to Drone Detection via Network Traffic Analysis},
  journal   = {Computer Networks},
  volume    = {168},
  pages     = {107044},
  year      = {2020}
}

@article{adel2024watch,
  author    = {A. Adel and T. Jan},
  title     = {Watch the Skies: A Study on Drone Attack Vectors, Forensic Approaches, and Persisting Security Challenges},
  journal   = {Future Internet},
  volume    = {16},
  number    = {7},
  pages     = {250},
  year      = {2024}
}

@inproceedings{more2025holystone,
  author    = {S. More and others},
  title     = {Comprehensive Security Assessment of Holy Stone Drones: Examining Attack Vectors},
  booktitle = {Proc. Int. Conf. Cyber Warfare and Security},
  volume    = {20},
  number    = {1},
  pages     = {574--583},
  year      = {2025},
  doi       = {10.34190/iccws.20.1.3286}
}

@mastersthesis{more2024thesis,
  author    = {S. A. More},
  title     = {Security Analysis of {HolyStone} Drones: Examining Attack Vectors and Data Extraction Techniques},
  school    = {Florida Institute of Technology},
  address   = {Melbourne, FL},
  year      = {2024}
}

@misc{sadhu2020fault,
  author    = {V. Sadhu and S. Zonouz and D. Pompili},
  title     = {On-board Deep-learning-based Unmanned Aerial Vehicle Fault Cause Detection and Identification},
  howpublished = {arXiv:2005.00336},
  year      = {2020}
}

@misc{taylor2021avis,
  author    = {M. Taylor and others},
  title     = {{Avis}: In-Situ Model Checking for Unmanned Aerial Vehicles},
  howpublished = {arXiv:2106.14959},
  year      = {2021}
}

@inproceedings{knox2022crazysim,
  author    = {D. Knox and others},
  title     = {{CrazySim}: A Software-in-the-Loop Simulator for the Crazyflie Nano Quadrotor},
  year      = {2022}
}

@inproceedings{llanes2024framework,
  author    = {S. Llanes and others},
  title     = {A Practical Framework for Multi-Agent Experiments in Aerial Robotics},
  year      = {2024}
}

@inproceedings{vasile2019breaking,
  author    = {S. Vasile and D. Oswald and T. Chothia},
  title     = {Breaking All the Things---A Systematic Survey of Firmware Extraction Techniques for {IoT} Devices},
  booktitle = {Smart Card Research and Advanced Applications ({CARDIS})},
  series    = {Lecture Notes in Computer Science},
  publisher = {Springer},
  pages     = {171--185},
  year      = {2019},
  doi       = {10.1007/978-3-030-15462-2\_12}
}

@article{ulhaq2023survey,
  author    = {S. Ul Haq and others},
  title     = {A Survey on {IoT} \& Embedded Device Firmware Security: Architecture, Extraction Techniques, and Vulnerability Analysis Frameworks},
  journal   = {Discover Internet of Things},
  volume    = {3},
  number    = {1},
  pages     = {17},
  year      = {2023},
  doi       = {10.1007/s43926-023-00045-2}
}

@mastersthesis{whipple2020survey,
  author    = {B. A. Whipple},
  title     = {A Survey of Firmware Analysis Techniques and Tools},
  school    = {University of Idaho},
  year      = {2020}
}

@inproceedings{tekeoglu2016testbed,
  author    = {A. Tekeoglu and A. S. Tosun},
  title     = {A Testbed for Security and Privacy Analysis of {IoT} Devices},
  booktitle = {Proc. {IEEE} Int. Conf. Consumer Electronics},
  year      = {2016}
}

@inproceedings{yasin2022review,
  author    = {A. Yasin and N. Jayapandian},
  title     = {A Review on Cyber Security Issues and Research Challenges in Internet of Things},
  booktitle = {Proc. 2nd Int. Conf. Electronics, Communication and Aerospace Technology},
  year      = {2022}
}

@misc{chen2022metaemu,
  author    = {Z. Chen and S. L. Thomas and F. D. Garcia},
  title     = {{MetaEmu}: An Architecture Agnostic Rehosting Framework for Automotive Firmware},
  howpublished = {arXiv:2208.03528},
  year      = {2022}
}

@inproceedings{kim2021firmcov,
  author    = {M. Kim and others},
  title     = {{FIRM-COV}: High-Coverage Greybox Fuzzing for {IoT} Firmware via Optimized Process Emulation},
  year      = {2021}
}

@inproceedings{lee2023embedded,
  author    = {J. Lee and others},
  title     = {Embedded Firmware Rehosting System Through Automatic Peripheral Modeling},
  year      = {2023}
}

@article{udeshi2024remaqe,
  author    = {M. Udeshi and others},
  title     = {{REMaQE}: Reverse Engineering Math Equations from Executables},
  journal   = {{ACM} Transactions on Cyber-Physical Systems},
  volume    = {8},
  number    = {4},
  pages     = {1--25},
  year      = {2024},
  doi       = {10.1145/3699674}
}

@misc{kim2024challenges,
  author    = {Y. Kim and K. Cho and S. Kim},
  title     = {Challenges in Drone Firmware Analyses and Its Solutions},
  howpublished = {arXiv:2312.16818},
  year      = {2024}
}

@article{agarwal2025secupilot,
  author    = {Y. Agarwal and V. Raghunathan},
  title     = {{SecuPilot}: A Security Coprocessor-Integrated Platform for Autonomous {UAV} Security},
  journal   = {{ACM} Transactions on Embedded Computing Systems},
  volume    = {24},
  number    = {5s},
  pages     = {1--25},
  year      = {2025},
  doi       = {10.1145/3762642}
}

@article{seo2023blockchain,
  author    = {J. W. Seo and others},
  title     = {Blockchain-Based Secure Firmware Update Using an {UAV}},
  journal   = {Electronics},
  volume    = {12},
  number    = {10},
  pages     = {2189},
  year      = {2023},
  doi       = {10.3390/electronics12102189}
}

@inproceedings{wang2024dvatar,
  author    = {Z. Wang and others},
  title     = {Dvatar: Simulating the Binary Firmware of Drones},
  year      = {2024}
}

@misc{emba2022,
  author    = {M. Messner and P. Eckmann},
  title     = {{EMBA} -- From Firmware to Exploit},
  howpublished = {Presented at Black Hat Europe, London, UK. [Online]. Available: \url{https://github.com/e-m-b-a/emba}},
  year      = {2022}
}

@misc{flashrom2024,
  author    = {{flashrom contributors}},
  title     = {flashrom: Open-Source Flash Programming Utility},
  howpublished = {[Online]. Available: \url{https://flashrom.org}},
  year      = {2024}
}

@misc{binwalk2010,
  author    = {C. Heffner},
  title     = {Binwalk: Firmware Analysis Tool},
  howpublished = {ReFirm Labs. [Online]. Available: \url{https://github.com/ReFirmLabs/binwalk}},
  year      = {2010}
}

@misc{openocd2024,
  author    = {{OpenOCD contributors}},
  title     = {Open On-Chip Debugger},
  howpublished = {[Online]. Available: \url{https://openocd.org}},
  year      = {2024}
}

@article{shannon1948mathematical,
  author    = {C. E. Shannon},
  title     = {A Mathematical Theory of Communication},
  journal   = {Bell System Technical Journal},
  volume    = {27},
  number    = {3},
  pages     = {379--423},
  year      = {1948},
  doi       = {10.1002/j.1538-7305.1948.tb01338.x}
}

@techreport{arm2009trustzone,
  author      = {{ARM Limited}},
  title       = {{ARM} Security Technology: Building a Secure System Using {TrustZone} Technology},
  institution = {ARM Limited},
  number      = {PRD29-GENC-009492C},
  year        = {2009},
  howpublished = {[Online]. Available: \url{https://developer.arm.com/documentation/PRD29-GENC-009492/c}}
}
